\documentclass[authoryear,12pt,3p]{jowarticle}

\usepackage[english]{babel}
\usepackage[utf8]{inputenc}
\usepackage{amsmath}
\usepackage{graphicx}
\usepackage[colorinlistoftodos]{todonotes}
\usepackage{natbib}

\makeatletter
\renewcommand\section{\@startsection{section}{1}{\z@}{-3.25ex plus -1ex minus -.2ex}{1.5ex plus .2ex}{\normalsize\bf}}
\renewcommand\subsection{\@startsection{subsection}{2}{\z@}{-3.25ex plus -1ex minus -.2ex}{1.5ex plus .2ex}{\normalsize\bf}}
\renewcommand\subsubsection{\@startsection{subsubsection}{3}{\z@}{-3.25ex plus -1ex minus -.2ex}{1.5ex plus .2ex}{\normalsize\bf}}
\makeatother

\begin{document}
\begin{frontmatter}
\title{Methods, Models, and the Evolution of Moral Psychology}
\author{Cailin O'Connor}\ead{cailino@uci.edu}
\date{\today}
\begin{abstract}
Why are we good?  Why are we bad?  Questions regarding the evolution of morality have spurred an astoundingly large interdisciplinary literature.  Some significant subset of this body of work addresses questions regarding our moral psychology: how did humans evolve the psychological properties which underpin our systems of ethics and morality?  Here I do three things.  First, I discuss some methodological issues, and defend particularly effective methods for addressing many research questions in this area.  Second, I give an in-depth example, describing how an explanation can be given for the evolution of guilt---one of the core moral emotions---using the methods advocated here. Last, I lay out which sorts of strategic scenarios generally are the ones that our moral psychology evolved to `solve', and thus which models are the most useful in further exploring this evolution. 
\end{abstract}

\end{frontmatter}

\section{Introduction}

Why are we good?  Why are we bad?  Questions regarding the evolution of morality have spurred an astoundingly large interdisciplinary literature with contributions from (at least) biology, sociology, anthropology, psychology, philosophy, and computer science.  Some significant subset of this body of work addresses questions regarding our moral psychology: how did humans evolve the \emph{psychological} properties which underpin our systems of ethics and morality?

Debates in this literature are ongoing, and often heated.  One reason for this is that there are a number of methodological problems researchers must tackle in addressing the evolution of moral psychology.  Human psychology evolved in the deep past, over the course of millions of years, and as a result of countless interactions between individuals.  The data available to us about these events are often sparse.  Further, it is often not even clear exactly what we are trying to account for when we attempt to explain the evolution of human moral psychology.  There are several interrelated issues here.  One issue arises from disagreements about what counts as our moral psychology in the first place.  Are we worried about traits we share with many animals, like prosocial feelings?  Or more high level traits like the ability to make moral judgments?  The second has to do with determining which aspects of human moral behavior are in fact the result of a biologically evolved psychology, and which are culturally learned.

Despite these, and other, issues, headway can be made.  But one must be careful.  The goal of this paper will not be to survey everything that has been done on the evolution of moral psychology---that would require a (large) book. (Those who wish to learn more might look at \citet{de1996good,sober1999unto,joyce2007evolution,churchland2018braintrust} or else \citet{james2010introduction} for a short overview.)  Here I will do three things.  First, I will discuss the methodological issues at hand at greater length, and defend what I think are particularly effective methods for addressing many research questions in this area.  In particular, I will argue that researchers need to use input from empirical literature and also evolutionary modeling techniques to triangulate well-supported theses regarding the evolution of various moral psychological traits. 

Second, I will give an in-depth example of this proposed methodology using one of the best modeling frameworks for this sort of exploration---game theory and evolutionary game theory.  I will describe how a partial explanation can be given for the evolution of guilt---one of the core moral emotions---using the methods suggested here. Last, I will carefully lay out more broadly which sorts of strategic scenarios are the ones that our moral psychology evolved to `solve', and thus which models are the most useful in further exploring this evolution.  Along the way, I will briefly discuss how work on each of these games can inform the evolution of moral psychology.

The paper will proceed as just described.  In section, \ref{sec:Methods}, I discuss methodological issues in the study of the evolution of moral psychology.  In section \ref{sec:Guilt}, I discuss the evolution of guilt, demonstrating how empirical work and evolutionary modeling complement each other in such an exploration. And in \ref{sec:Models}, I describe several types of games---including the famous prisoner's dilemma, the stag hunt, bargaining games, and coordination games---to make clear what challenges and opportunities of social life led to the emergence of moral behavior and moral psychology. 

\section{Methods}
\label{sec:Methods}

As mentioned in the introduction, there are a number of methodological issues plaguing the study of the evolution of moral psychology.  Here I outline them in more detail.  As we will see, these issues are sometimes intertwined.

Many of the historical sciences face problems related to sparse data.  In paleontology, scientists may have access to a single bone, and from that evidence alone attempt to infer features of an entire species.\footnote{For more on methodology in the historical sciences see, for example, \citet{chapman2016evidential,currie2016hot,currie2018rock}.}  In the case of human behavioral evolution there are various types of historical data available: bone and tool remains, footprints, cave drawings etc.  Scientists also draw on data from modern primates, hunter-gatherer societies, and human psychology in developing coherent evolutionary narratives regarding our psychology.

As \citet{longino1983body} painstakingly outline, sometimes historical scientists can draw fairly dependable inferences based on their data.  Other times, especially when it comes to the evolution of human psychology, these inferences will have to depend on premises that are themselves shaky or unconfirmed.\footnote{Another way to diagnose the issue here is that in some cases the historical evidence more severely underdetermines what hypotheses are drawn \citep{sep-scientific-underdetermination}.}  One central worry about much work in evolutionary psychology---a discipline aimed specifically at explaining human behavior from an evolutionary standpoint---is that the `gaps' in these inferences can leave room for current cultural beliefs and biases to creep in.\footnote{For an example, see the debate between evolutionary psychologists \citep{tooby1992psychological,buss1995evolutionary} and social structural theorists \citep{eagly1999origins} on the origins of gender differences in psychology.} 

A way to further constrain these evolutionary narratives, and thus avoid some of these worries, is to use evolutionary modeling techniques.  One branch of work along these lines comes from biology where researchers have used population genetic models to assess, for example, the conditions under which altruistic behaviors can evolve.\footnote{The most influential work along these lines begins with \citet{hamilton1964genetical} on kin selection.  See \citet{sep-altruism-biological} for an overview.}  Another related branch of work using evolutionary game theory starts with static representations of strategic social interactions and adds dynamical aspects to the model to see how behaviors in these scenarios evolve.  More on this in the next section.

These sorts of models have been very successful, as we shall see, in explaining the evolution of moral behaviors such as altruism, cooperation, and apology.  For example, they have been used to disconfirm evolutionary narratives about these behaviors that seemed coherent, thus improving on what we can do with sparse data and human reasoning alone.\footnote{A famous example comes from John Maynard-Smith's modeling work \citep{smith1964group} which played a central role in refuting work on the group selection of altruism \citep{wynne1962animal}.} For this reason, I advocate here for the importance of modeling in the study of the evolution of moral psychology.  But the use of these models raises another methodological issue.  As mentioned, they address the evolution of behavior, not psychology.  The reason is that, when it comes to evolution, behavior is where the rubber meets the road.  The internal psychological organization of an organism only matters to its fitness inasmuch as it influences how the organism behaves.  Of course, psychology and behavior are tightly connected since psychology evolves for the purpose of shaping effective behavior.  But if we start with a model for a successful, fitness enhancing behavior, it takes an extra step to argue that such a model explains the evolution of a psychological trait.\footnote{Another traditional worry about the use of evolutionary models to understand behavior has to do with employing simplified representations to understand a complex world.  Philosophers of modeling have worked extensively on this topic.  (\citet{weisberg2012simulation} surveys much of this work.)  I do not address this worry here.} 

To see why, let's consider an example.  Suppose we look at a model of the prisoner's dilemma and observe that under some set of conditions altruistic behavior is selected for.  If we want to connect this observation to human psychology, we can argue that under these same conditions psychological traits that promote altruism will likewise be selected for.  But the model does not tell us what those psychological traits are---anything that causes altruistic behavior might do.  To make the connection to human psychology, then, we need to know what the candidate psychological traits responsible for altruism are ahead of time.  This is the sense in which empirical data and models must be combined in studying the evolution of human psychology.  One has to know empirically what psychological traits humans have, and what behaviors associate with those traits, in order to use models to explain them.  Note that this makes it quite difficult to predict what sorts of psychological traits \emph{will} evolve given a selective environment, even if we can generate good evidence about why certain psychological traits \emph{did} evolve.

So, to use evolutionary models to study the evolution of human moral psychology, we need to know what moral psychology we are trying to explain.  But this is not always simple.  A first issue along these lines has to do with figuring out which aspects of moral psychology are biologically evolved.  This can be tricky because much of human moral behavior is culturally shaped.  Moral systems vary greatly across cultures, so we know there is no robustly hardwired biology that fully determines our moral psychology.  These systems emerge on a cultural evolutionary timescale, and individual psychology is shaped by these cultural systems in multiple ways during the course of a lifetime.  

On the other hand, it is also clear that much of our moral behavior is shaped by evolved biological tendencies.  Some moral `non-nativists' have argued that this is not so.  One view is that human morality emerged as a side-effect of human-level rationality \citep{ayala1987biological}.  More common is the view that morality is entirely, or almost entirely, learned, not evolved \citep{prinz2008morality}.  Empirically, these views seem untenable. \citet{joyce2007evolution} surveys literatures showing that, for example, 1) children show early, strong empathetic tendencies, 2) children develop a remarkably early understanding of norms, including the difference between conventional and moral norms, 3) primates and other animals display psychological traits connected to human morality, like prosocial pleasure and prosocial distress, 4) it is well established that emotions play a key role in moral decision making, and 5) sociopaths display perfectly good reasoning and learning skills, but nonetheless fail to engage in normal moral behavior as a result of emotional deficits.  So to sum up, the evidence is that moral psychology is produced by three interacting processes on different timescales, 1) biological evolution, 2) cultural evolution, and 3) individual learning.\footnote{For more on some of the ways these processes might interact see \citet{nichols2004sentimental, sripada2008nativism}.}

How do we pick apart just those aspects that are well-explained by biological evolution?  There is no general answer.  Researchers have to argue it out on a case-to-case basis using the empirical evidence available.  In many cases a full explanation of some moral psychological feature will have to draw on all three processes.  For an emotion like guilt, for instance, we know that its production is highly culturally dependent, i.e., responsive to culturally evolved norms.  On the other hand, it has clear biological components since sociopaths seem to lack normal capacities to feel guilt \citep{blair1995cognitive,blair2005psychopath}.  Furthermore, various aspects of moral education play a role in the prevalence and strength of guilty feelings \citep{benedict2005chrysanthemum}.  And, to add one more complication, in this case cultural evolutionary processes and biological evolutionary processes likely feedback on each other in a gene-culture co-evolutionary process \citep{gintis2003hitchhiker}.  For instance, culturally evolved moral norms for punishment create a selective scenario where guilt is more likely to emerge \citep{o2016evolution}.  At the same time, biological tendencies towards guilt shape which moral systems are culturally viable.  The Baldwin effect from biology occurs when a learned behavior eventually becomes more and more innate.\footnote{This was first proposed by \citet{baldwin1896new}.}  There may well have been various moral psychological traits that emerged on a cultural timescale, but have subsequently been stabilized by selection in this way.

In other words, it is complicated.  These particular complications suggest another step for the methodology proposed above.  One should see what behaviors are produced by a moral psychological trait, and use modeling to understand what scenarios support selection of those behaviors.  Once this is done, though, a further question to ask, which must necessarily depend on empirical data is: are those selection pressures culturally created or not?  Is the solution to those pressures a cultural system that shapes our psychology or a biological one? (Of course, this need not be a clean distinction, and, in fact, will probably not be.)  Answering these questions may help researchers refine their models to produce more satisfactory explanations of moral psychological traits.  (In addition, as we will see in section \ref{sec:Guilt}, returning to empirical data with plausible modeling results can often lead to refinements of other sorts as well.)

Another methodological issue, again regarding which psychological traits are candidates for evolutionary explanation, has to do with determining which traits are really the `moral' ones.  There are two sub-problems here.  The first has to do with general problems regarding the identification and delineation of psychological traits.  Which are really distinct from the others?  For instance, emotions researchers mostly agree about the existence of the `big five' emotions, but there is huge disagreement about how many other emotions there are, what these are, and how they are related.\footnote{Compare, for example, the eight emotions identified in \citet{plutchik1991emotions}, and organized as opposites, with the 27 emotions identified by \citet{Cowen201702247}.}  These debates get muddied by cultural influences on the development of moral psychological traits.  For example, debates about the definitions of guilt and shame have been confused by the significant cultural variation in the production of these emotions \citep{wong2007cultural}.\footnote{\citet{benedict2005chrysanthemum} gave an influential, early analysis of these cultural differences.}

The second sub-problem, which has mostly arisen in philosophy, has to do with identifying which behaviors genuinely count as `moral'.  Should we be focusing mostly on prosocial instincts?  Emotions?  Or higher level abilities unique to humans?  Theory of mind?  The ability to make moral judgments? I follow \citet{churchland2018braintrust} in being generally uninterested in trying to disambiguate truly moral behavior from behavior in the animal kingdom with moral character.\footnote{Furthermore \citet{Stich2016} compellingly argues that this boundary question does not have a good answer.}  The most compelling, naturalistic accounts admit that our moral psychology involves many features shared with other animals, and also a number of features, like theory of mind and language use, that few species (or just humans) exhibit.  As such, there are a large variety of diverse moral psychological traits one might want to make sense of in the light of evolution.

Before completing this section, I think it will be helpful to develop an (incomplete) list of aspects of human psychology that contribute to moral behavior and thus might be appropriate targets of evolutionary explanations of moral psychology.  Note that this list will include concepts that overlap, and others that are ill-defined.   Many of the things on this list can be tackled to some degree of success using the kind of methodology I advocate here, though.  (Indeed, many of them already have been addressed from an evolutionary point of view, but it is beyond the scope of this paper to fully survey this literature.)  

There are positive prosocial emotions, instincts, and feelings including social trust, love, empathy, sympathy, compassion, and gratitude.  There are also negative prosocial feelings and emotions such as guilt, shame, and embarrassment.  Empathy can also lead to prosocial distress, for instance if one sees a loved one in pain.  In addition, we have a set of emotions that contribute to punishment and norm guidance: moral anger, urges for retribution, contempt, and disgust. 

Besides the emotional side of things, \citet{joyce2007evolution} argues that the ability to judge something as morally wrong is key to human morality.  In other words, if we did not have the capacity to think thoughts like, `Jill ought not have done that' or `Brian deserves punishment for breaking rule X', we simply would not have full human morality. As Joyce argues, language is a prerequisite for this sort of judgment.\footnote{While language might be needed for judgment of the sort Joyce has in mind, others have argued that certain moral abilities preceded and coevoled with language.  Some significant level of cooperation and trust, for instance, may have been necessary to generate the social structure that selected for language \citep{richerson2010possibly, sterelny2012language}.}  He further argues that this judgment is a trait that has been biologically selected for, rather than emerging from other cognitive abilities.  (On his account the benefits of this trait come from the fact that moral judgments solidify prosocial behaviors.)

Another such trait is moral projection.  We often feel that moral norms like `do not kill innocent people' are not just wrong for us, or wrong in our culture, but are simply wrong in a freestanding, objective way.\footnote{For instance, in one study Amish school children were asked `if God said it was OK to work on Sunday would it be?'  They agreed.  When stealing was substituted for `working on Sunday', they largely did not agree \citep{nucci1985children,nucci2001education}.} \citet{stanford2018difference} develops an evolutionary narrative where this sort of trait benefits individuals by improving moral coordination in that individuals who judge X right are also incentivized to perform X.  For \citet{joyce2007evolution} the evolution of moral projection was a key step in our ability to make genuinely moral judgments.

\citet{churchland2018braintrust} and others have argued that theory of mind---the ability to understand others as having minds like our own---is a key feature of our moral psychology.  Without it, we cannot judge intent, and thus cannot make standard moral judgments.\footnote{It should be noted that the role intent plays in moral judgment shows cross-cultural variation.  Some cultures seem to ignore intent in making moral judgment \citep{barrett2016small}.}  In addition, theory of mind plays a role in empathy by allowing us to conceptualize feelings in others that we do not feel ourselves.

Various scholars have noted that humans are fascinated by gossip. \citet{emler1990social} argues that humans spend 60-70\% of conversations on gossip and `reputation management'.  This trait is important to our moral systems, where individuals use knowledge about past moral behavior to choose interactive partners \citep{joyce2007evolution}.  It also plays a role in punishment and reciprocation.\footnote{Some have even argued that gossip is so important to humans, it drove the evolution of human language \citep{aiello1993neocortex}.}

Some have argued that humans, and maybe even non-human primates have an innate aversion to inequity, or a `taste for fairness' \citep{fehr1999theory}.  If so, this is likely connected to the prevalence of norms for fairness and justice across human groups. In a reflection of the methodological worries raised above, there has been debate in economics about whether human behavior is better explained by inequity aversion (which presumably is at least partially a biological trait) or by social norms for equity (which emerge on the cultural timescale) \citep{fehr1999theory,binmore2010,fehr2010inequity}.  Most seem to agree that there is at least some innate inequity aversion of this sort. 

Norms are rules in human groups which may vary cross-culturally, and which determine what behaviors `ought' or `ought not' be performed.  While not all norms have a moral character, there seem to be psychological traits dedicated to norm governed behavior which play a strong role in human moral systems.  As \citet{sripada2005framework} outline, one of these traits involves internalizing norms, which is a process whereby individuals come to believe they should follow them regardless of expectations for sanction.  This psychological trait plays a key role in moral systems \citep{joyce2007evolution,stanford2018difference}.

Associated with norm governance and norm psychology are learning behaviors that allow groups of humans to adopt group-specific norms and conventions.  First is the simple fact that humans have evolved to be extraordinary social imitators.  Beyond this, humans show biases towards learning common behaviors, as well as behaviors displayed by particularly prominent or successful group members \citep{boyd1988culture}.  These learning abilities are key to the stability of our moral norms.
 
I will mention one more psychological trait which is certainly important in the production of our moral behavior, even if we might not want it to be.  Humans show strong psychological tendencies towards in-group bias, i.e., treating in-group members better than out-group members.\footnote{Evidence for this bias comes from experiments in `minimal group paradigm'.  These experiments involve the formation of often arbitrary groups, i.e., by coin flip, and find that participants nonetheless use group structure to determine prosocial behavior \citep{tajfel1970experiments}.}  This tendency likely evolved to regulate human prosocial behavior.\footnote{I will not describe the evolutionary models most germane to understanding this aspect of human behavior, but see, for example \citet{boyd2005origin,oconnorineq2018}.}

\section{Modeling Guilt}
\label{sec:Guilt}

The methodology I am advocating for, as noted above, involves 1) choosing some aspect of human moral psychology, 2) carefully investigating which behaviors it causes under what conditions, 3) using evolutionary models to explain the psychological trait by explaining the evolution of the behaviors it causes, and 4) returning to empirical data to refine the model, and further develop the narrative one can draw from it.  At this point, I move on to the second part of the paper, which involves giving an in-depth example of how this methodology can work.  In particular, I describe how empirical and modeling work complement each other in an investigation of the evolution of guilt.  Before doing so, though, I introduce basics from the mathematical frameworks used in this investigation---game theory and evolutionary game theory.

\subsection{Game theory and Evolutionary Game Theory}
\label{sec:EGT}

\emph{Game theory} is a branch of mathematics used to model \emph{strategic} behavior---behavior that involves multiple actors where each individual cares about what the others do.\footnote{This framework originated with the work of \citet{morgenstern1953theory}.}  A \emph{game} is a model of some such situation involving three elements: \emph{players}, or who is involved, \emph{strategies}, what actions they can take, and \emph{payoffs}, or what each player gets for each set of strategies played.\footnote{Usually games also define a fourth element, \emph{information}, or what each player knows about the game.  I ignore this element here since it is not usually relevant to evolutionary analyses.}

Traditional approaches to game theory model various strategic scenarios and use these models to predict or explain human behavior via assumptions about rationality.  For example, by showing that taking behavior X will always yield a lower payoff than behavior Y in some game, a theorist can generate the prediction that humans will not take behavior X in analogous scenarios.  Why use game theory here?  Moral behavior occurs in the social realm, i.e., it evolves as a solution to social problems.  Games are just those scenarios where actors interact socially in ways that matter to their fitness, and so just those scenarios morality emerged to regulate.

\emph{Evolutionary game theory}, first developed in biology, and then imported into the social sciences, asks: what sorts of strategic behaviors are likely to evolve among group of agents playing games?\footnote{This field originated with \citet{smith1973logic}, though precursor work occurred in economics.}  These models take a population in some strategic scenario and add what are called \emph{dynamics}, or rules for how the population will evolve.  The most widely used class of dynamics in evolutionary game theory assumes that strategies that yield higher payoffs will expand in a population, while those that yield lower payoffs will contract.  It should be clear why evolutionary game theory is a useful framework here.  It is intended to model just those sorts of scenarios most likely to inform moral psychology: situations where individuals evolve social behavior. 

\subsection{Guilt}

Guilt is a puzzling emotion in many ways.  Some emotions, like fear, have clear fitness benefits (avoiding danger).  Guilt is associated with a number of behaviors that seem straightforwardly detrimental, such as altruistic behavior, acceptance or seeking of punishment, making costly reparations, and even self punishment.  For this reason, and because of its relevance to ethical behavior, a number of philosophers have gotten interested in the evolution of guilt \citep{joyce2007evolution,deem2016guilt,DeemRamsey2,o2016evolution,rosenstock2018s}.  The goal here will be to demonstrate how the method described earlier in the chapter can help explain some aspects of the evolution of guilt.  I will draw on previous work from \citet{o2016evolution,rosenstock2018s}.  

Let's start, as suggested, by looking at empirical work to identify what sorts of strategic behaviors are associated with feelings of guilt.  These can be grouped into three rough categories.  First, guilt seems to prevent social transgression and is thus associated with prosocial behaviors such as cooperation and altruism \citep{tangney1996shame,regan1972voluntary,ketelaar2006role}.  Second, guilt leads to a suite of behaviors after social transgression, such as apology, acceptance of punishment, self punishment, and costly reparation \citep{silfver2007coping,ohtsubo2009sincere,nelissen2009guilt}.  Last, expressions of guilt seem to influence punishing and judging responses by other group members.  In particular, individuals who express feelings of guilt and remorse are more likely to be judged guilty (in the sense that they did the deed), but also more likely to be forgiven and punished less harshly \citep{eisenberg1997but,gold2000remorse,fischbacher2013acceptance}.

Let us focus, for the purposes of this chapter, on the second suite of behaviors associated with guilt---those that occur after transgression.  These, as pointed out, lead to immediate fitness costs and so seem particularly in need of an evolutionary explanation.  It is clear that these sorts of reparative behaviors play a role in avoiding ostracism after transgression \citep{joyce2007evolution}.  In order to better understand how this might work, let us now turn to an evolutionary model.

In the next section, I will describe at much greater length the \emph{prisoner's dilemma} and discuss how it can be used in understanding the evolution of moral emotions generally.  For now, it will serve to know that this is a model where actors have two options---to behave altruistically by playing `cooperate' or to behave selfishly by playing `defect'.  There is always a payoff incentive to choose defection, which, as we will see, raises the question of why altruism has so often evolved.  In the \emph{iterated} version of the prisoner's dilemma actors play again and again over the course of many round.  This makes it a particularly good model for exploring behaviors associated with reparation and apology---actors engaged in a long enough interaction to have time to potentially rend and repair their relationship.  (It also is a good model of early human interaction, since the group structures of early humans led to repeated interactions with group members.)

In particular, let's consider a version of this game where actors sometimes make mistakes.  An actor might usually behave altruistically by playing cooperate, but sometimes defect either accidentally, or due to exigent circumstances.  This might represent the sort of situation where a usually prosocial individual transgresses against ethical norms.  Let's suppose further that actors tend to use reciprocating strategies.  We will discuss these further in the next section, but the basic idea is to meet cooperative behavior with cooperation and selfish behavior with defection.  Accidental defection of the sort just described causes problems for these sorts of reciprocating strategies.  In particular, reciprocators can get stuck in cycles of retribution where they continue to defect on each other, despite have underlying altruistic tendencies.\footnote{There are various ways to solve this problem \citep{axelrod1981evolution,nowak1993strategy}.  As we will see, apology is one of these.}  Two altruistically-inclined individuals can lose the benefits of mutual altruism as a result of an error plus reciprocation.  

One way to get around this issue is through apology.  Models have shown that strategies where actors apologize after defection, and accept the apologies of others, can evolve since they avoid the payoff loss associated with mutual negative reciprocation.\footnote{See \citet{okamoto2000evolution,ohtsubo2009sincere,ho2012apologies,pereira2013so,pereira2016guilt,o2016evolution,rosenstock2018s}.  In \citet{pereira2016guilt,o2016evolution,rosenstock2018s} authors connect these models to the evolution of guilt.  In addition, \citet{pereira2017evolutionary,pereira2017social} take a different tack in showing how guilt can evolve by causing self-punishment.}  In showing how apology and reparation can evolve, these models can also show how guilt might be selected for in order to mediate apology and reparation.

One of the most important take-aways from these models is that this sort of apology can only work if it is hard to fake, costly, or both.  The reason is that unless there is some mechanism guaranteeing the apology, fake apologizers can take advantage of trusting group members.  If some individuals forgive and forget after being defected on, fakers can keep on saying `I'm sorry' and defecting in the next round.  This prevents apology from evolving to promote altruism.

Let us talk through these ways of stabilizing apology, and discuss how they might connect up with the evolution of guilt.  \citet{frank1988passions}, in very influential work, suggests that moral emotions evolved for a signaling purpose.  Guilt leads to altruistic behavior, and is also hard to fake (because it is an emotion).  Thus guilt can allow individuals to choose altruistic partners and avoid defectors.  Something similar might work for apology too.  If we can tell which individuals are offering sincere apologies, because we can simply see that they feel guilty, then we can only forgive those who have real, altruistic tendencies.  This is the sense in which apology can evolve if it is hard to fake \citep{o2016evolution}.

If we return to the empirical literature, though, we see that this model does not quite fit the bill with respect to explaining the evolution of guilt.  In particular, unlike some emotions (like anger and fear), there are no stereotypical body and facial postures associated with guilt.  This means that others cannot identify a guilty individual just by looking.  In other words, guilt does not display the features we would expect from an emotion that evolved for the purposes of signaling altruistic intent \citep{deem2016guilt}.

What about costly apology?  In models of apology and reparations, costs stymie fakers by making apologies not worth their while.  This is because apology yields different benefits for those who intend to defect and those who intend to re-enter a long, cooperative engagement.  Imagine you are the latter type of individual. You are being shunned from a group for a social transgression.  If you are willing to issue a costly apology by accepting punishment, punishing yourself, or paying the one you transgressed against, your group members will start cooperating with you again.  It should be well worth your while to pay even a large cost to re-enter the fold.  Doing so earns you a lifetime of receiving the benefits of mutual altruism. Now imagine you are a defector and plan to transgress immediately after apologizing.  Paying a cost to apologize will generally not be worth your while.  As soon as you defect, you'll be on the outs again, and have to pay another cost to find a cooperative partner.

The success of costs in facilitating the evolution of apology seems to tell us something important about the evolution of guilt.  There is a reason that guilt leads to a number of behaviors that are individually costly, which is that without these costs it would not be an effective emotion for promoting apology and reparation.

We can again turn to the empirical literature to tune this model further.  Remember that expressions of guilt seem to decrease punishing behaviors by group members. Although it does not make sense to posit that guilt evolved solely for a signaling purpose, this literature tells us that it does seem to play some signaling role in apology.  In \citet{rosenstock2018s} we explore in more detail why this might be.  In particular, we go back to the models and show that the high costs necessary to guarantee apology have a downside.  As costs, they directly decrease the fitness of those who feel guilt and issue costly apologies.  This means that even when guilt and apology are evolutionarily viable as a result of these costs, they may be unlikely to evolve.\footnote{We show that the basins of attraction for guilt-prone strategies that are generally cooperative, retributive, apologetic, and trusting of apologies are relatively small when costs are high.}  However, if expressions of guilt are even a bit trustworthy, this significantly decreases the necessary costs, and increases the likelihood that apology and guilt can evolve.\footnote{This combination of costs and honest signals was inspired by \citet{huttegger2015handicap} who show how biological signals that are somewhat hard to fake can reduce the costs necessary to ensure they are trustworthy.}  In other words, these slightly more complicated models show how guilt and expressions of guilt might emerge to regulate costly apology, while reducing the necessary costs as much as possible.

This work provides only a partial explanation of guilt.  Notice that the models invoked are evolutionary models, but they are not gene-culture coevolutionary models.  And, as mentioned, guilt likely is the result of gene-culture coevolution.  Furthermore, they do not explicitly represent or account for the role of culture in the production of guilt, or for the cultural variability in what sorts of transgression lead to apology and reparation.  (It is not always failures of altruism in the real world.)  As suggested, a good next step is to return to the empirical literature to hone the partial explanation developed here.

\section{Models}
\label{sec:Models}

We have now seen how empirical investigation and evolutionary modeling can be used in concert to elucidate the evolution of a particular moral psychological trait.  The goal of this section is to sketch out the directions in which such a methodology might be further applied.  I overview the main sets of models used to explain the evolution of moral behavior.  In other words, I lay out here the models that have the greatest potential to tell us something about the evolution of the panoply of moral psychological traits listed in section \ref{sec:Methods}.

\subsection{Altruism and the Prisoner's Dilemma}

The prisoner's dilemma, mentioned above, is probably the most famous, and certainly the most widely studied, game in game theory.  We have seen how it might play a role in modeling the evolution of guilt.  Let's now fill in the details and try to understand more generally how it can be used to model the evolution of moral psychology. 

The motivating story is that two prisoners are being asked to rat each other out.  If they both stay silent they each get a short jail sentence.  If they both rat, they each get a long one.  If only one rats on the other, who stays silent, the rat gets to go free, while the other one serves an even longer sentence.  

\begin{figure}
\centering
\includegraphics[width=.6\textwidth]{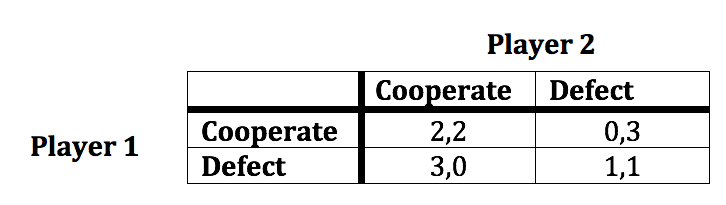}
\caption{A payoff table of the prisoner's dilemma.  There are two players, each of whom choose to cooperate or defect.  Payoffs are listed with player 1 first.
\label{fig:PD}}
\end{figure}

To turn this scenario into a game we must define the formal elements listed in section \ref{sec:EGT}.  The players here are the two prisoners.  Their strategies, to rat or stay silent, are usually labeled `defect' and `cooperate', as noted earlier.  Their payoffs in terms of \emph{utility} gained for each outcome described, are listed in figure \ref{fig:PD}. Each entry in the table shows the payoffs for some combination of strategies with player 1 listed first.  Utility here is meant to track preference or benefit for each player.  The numbers are chosen arbitrarily, but the game will be a prisoner's dilemma as long as they retain their ordering.

We can now see why this game is referred to as a `dilemma', and why it has generated such a significant literature.  Both players prefer mutual cooperation over mutual defection, because the first yields payoffs of 2 and the second 1.  But regardless of what the other player is doing, each player prefers to defect rather than cooperate.  If your opponent cooperates, you receive a 3 for defection and only a 2 for cooperation.  If your opponent cooperates you receive a 1 for defection and only 0 for cooperation.  This means that defect-defect is the only \emph{Nash equilibrium} of the game---the only pairing of strategies where neither player wants to change what they are doing.  (This is a central predictive concept in game theory and evolutionary game theory.  Because no one wants to deviate, Nash equilibria tend to be stable, and also tend to emerge in evolutionary models.)  Importantly, this payoff structure also means that the `cooperate' strategy can represent altruistic behavior.  Taking this strategy always involves lowering your own payoff while simultaneously increasing the payoff of your partner.  Thus the prisoner's dilemma serves as a model for a wide swath of human moral behavior, including those we discussed in the last section on guilt.

At first glance, the prisoner's dilemma model seems to predict non-altruistic behavior.  It is always better to defect than to cooperate.  But, of course, humans engage in altruistic behavior all the time.  (To give one cute example, I just watched an internet video where a group of men spent over an hour cutting a humpback whale free from a fishing net.)  How do we explain this discrepancy?  Evolutionary game theory provides one avenue.  The suggestion is that a population of biological agents might evolve altruistic traits, even though on the face of it altruism seems irrational.  And, regarding moral psychology, such an agent might thus come under selection pressure for psychological traits that lead to altruistic behavior.  

The key to evolving altruism in a scenario where actors face a prisoner's dilemma is \emph{correlated interaction} between the strategies.  Whenever cooperators meet cooperators and defectors meet defectors often enough, cooperation can win out.  We can think of correlation as changing the payoff table in figure \ref{fig:PD} to one where only the top left and bottom right entries are available.  And in such a case, cooperation does strictly better.  

So what are the mechanisms that lead to correlated interaction of this sort?  There are a number that have been identified.\footnote{See \citet{nowak2006five} for a description of five major mechanisms that have played significant roles in the literature on the evolution of altruism: kin selection, group selection, direct reciprocity, indirect reciprocity, and network reciprocity.}  I will focus here on kin selection, direct and indirect reciprocity, and punishment as the ones most pertinent to the evolution of moral psychology.

Kin selection explains the emergence of altruistic behavior across the animal kingdom.\footnote{The theory of kin selection was introduced by \citet{hamilton1964genetical}, and has been developed extensively since then, as in \citet{dawkins2006selfish,michod1982theory,grafen1984natural}.}  Kin, of course, are more likely to share genetics.  If altruists interact with their own kin, then the benefits they give tend to fall on other altruists.  This can make altruistic genes more successful than selfish genes.  This also helps explain why throughout the animal kingdom, organisms are most likely to give care to their own young, siblings, and family groups.  Most theorists think that the first origins of moral psychology---positive, caring feelings directed towards kin, and distress at pain or separation from kin---were selected for via kin selection in our ancestors \citep{joyce2007evolution,james2010introduction, churchland2018braintrust}.

Reciprocal altruism is common in human moral systems, and also present in a few species of non-human animal.\footnote{Famously, in the vampire bat \citep{wilkinson1984reciprocal}.} The basic idea behind reciprocity is that players tend to cooperate with those who have cooperated in the past, and defect against those who have defected.  The net outcome is that cooperators end up fitter, despite their altruistic behavior, because they are treated altruistically.  As \citet{trivers1971evolution} first showed in biology, this means that reciprocally altruistic behavior can evolve.  

Direct reciprocity occurs when individuals choose their strategies based on how a partner has treated them in the past.  This sort of behavior is often modeled using the \emph{iterated prisoner's dilemma} mentioned in the last section, where two actors play the prisoner's dilemma again and again.  One example of a reciprocating strategy made famous by \citet{axelrod1981evolution} is tit-for-tat (TFT), where each player starts cooperative and then takes the strategy that their opponent did in the last round. This strategy does well against defectors, cooperators, and itself.  And (though things are a bit complicated), direct reciprocity in general, and the tit-for-tat strategy itself can evolve \citep{trivers1971evolution, axelrod1981evolution,nowak2006evolutionary}.  Another such strategy is the grim trigger, where actors cooperate until a partner defects, and then defect forever.\footnote{See \citet{sachs2004evolution} for a good overview of the evolution of reciprocal altruism.}  Notice that these evolutionary models lend themselves to explaining the emergence of moral psychology surrounding retribution, contempt, and moral anger, and the connection between this psychology and altruism in human groups.

Indirect reciprocity involves reciprocating on the basis of transmitted information.  If John defects against me today, and I defect against him tomorrow, this is direct reciprocity.  If John defects against me today, and I report it to you, and you defect against him tomorrow, this is indirect reciprocity.  Reputation tracking of this sort can help promote the emergence of altruistic strategies \citep{alexander2017biology,nowak1998evolution}.  This sort of model, notice, lends itself to explaining the moral psychology behind gossip and reputation tracking.

The last thing to discuss here is punishment, though this route to the evolution of altruism gets a bit complicated.  In some sense reciprocation is a type of punishment, but we might also consider actors who actively lower the payoffs of defectors, rather than just defecting on them in the future.  The idea here is that if defectors are regularly punished for defecting, altruists can get a leg up.  We do, of course, see moralistic punishment, including for altruism failures, across human societies \citep{fehr2002altruistic,henrich2006costly}.  And we know that such punishment can help maintain altruistic behavior \citep{yamagishi1986provision,ostrom1992covenants}.  But the evolution of punishing behavior is complicated by the fact that punishing others is always at least a little costly, leading to what is termed the `second order free rider problem'.  Why would I sacrifice my fitness to punish a violator?  A number of solutions have been proposed, though discussing them is beyond the scope of this paper.\footnote{For example, see \citet{panchanathan2004indirect}. See \citet{frank2003repression} as well.} Note that these models may serve to help us further understand the moral psychology behind punishment---moral anger, indignation, etc.

To sum up, the prisoner's dilemma helps us frame one of the most significant social problems facing early humans---why be altruistic when selfishness pays off?  It also helps us answer one of the most pressing puzzles in the evolution of moral psychology---why are humans so psychologically inclined towards altruism?  As we saw, modeling work on the evolution of altruism is actually relevant to many aspects of moral psychology including prosocial instincts and emotions, as well as the psychology of retribution and punishment.

\subsubsection{Public Goods, Trust, and Punishment}

Before continuing, I would like to briefly mention another game that shares some character with the prisoner's dilemma.  In \emph{public goods games}, actors have the option to contribute some amount of resource to a public good.  This could represent a group of individuals clearing a meadow for public grazing, or building a town hall.  The total amount contributed is multiplied by some factor, and then equally divided among the contributors.  This captures the idea that there is a benefit to joint production---a group of people together can produce marginally more than they could alone.  It also captures the idea that in any such situation there is a temptation to shirk by not contributing, but still enjoying the fruits of others' labor.  The Nash equilibrium of this game is that each player contributes \$0, even though they all would do much better by each contributing the maximum (or anything more than 0).  For this reason, public goods games present another sort of social dilemma, where giving represents altruism in that any increase of personal contribution increases the payoffs of group members at the expense of one's own payoff.

This game has been widely studied experimentally, and punishment and reward have been shown to improve contributions \citep{fehr2000cooperation}.  From an evolutionary perspective, punishment, reward, and reputation tracking can lead to the evolution of altruisic behavior in public goods games \citep{hauert2010replicator}.\footnote{See \citet{santos2008social} for more on how diversity of population traits influences these processes.}  Like various evolutionary models of the prisoner's dilemma, these models may be able to tell us something about moral anger, retribution, gratitude, and the moral psychology behind altruism.  In addition, this model has been taken as informative of the evolution of social trust \citep{churchland2018braintrust}.

\subsection{Cooperation, Coordination, and the Stag Hunt}

The prisoner's dilemma gets a lot of attention, but altruism is not the only prosocial behavior that has emerged in human groups.  We now investigate another type of strategic scenario often faced by humans---where cooperation or joint action is mutually beneficial, but risky.

\begin{figure}
\centering
\includegraphics[width=.7\textwidth]{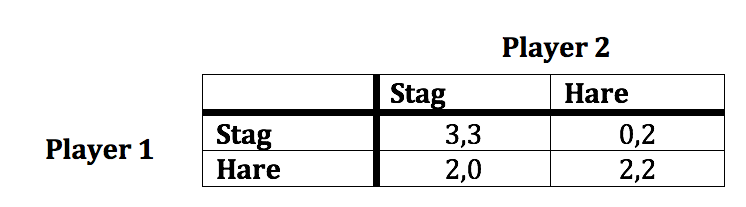}
\caption{A payoff table of the stag hunt.  There are two players, each of whom choose to hunt stag or hare.  Payoffs are listed with player 1 first.
\label{fig:staghunt}}
\end{figure}

This sort of scenario is typically modeled by a game called the \emph{stag hunt}.  The motivating story is that two hunters have the option to either hunt stag or hare.  If they hunt hare, they each gather a dependable, small amount of meat.  If they hunt stag, they have the opportunity to gather much more by cooperating---half a stag each.  But they will only catch the stag if both work together.  In other words, there is a risk to cooperating because one's partner might decide to work alone.\footnote{This motivating story is from \citet{rousseau1984discourse}.}  Figure \ref{fig:staghunt} shows this game.  The payoffs are 3 for joint stag hunting, 2 for hunting hare, and 0 for hunting stag when your partner hunts hare.  In influential work, \citet{skyrms2004stag} uses this game as a simple representation of the social contract---hare hunting represents a state of nature, and stag hunting a social organization.  In general, the game can capture many scenarios where by working together humans can generate surplus resources, but doing so opens them up to risk.

Unlike the prisoner's dilemma, this game has two Nash equilibria: the strategy sets where both actors hunt stag, or both hunt hare.  In other words, either a cooperative, mutually-beneficial outcome, or an uncooperative outcome is possible from an evolutionary standpoint.  From a naive point of view it might seem obvious which should evolve---stag hunting, right?  It yields a greater payoff.

Things are not so straightforward.  This is because stag hunting, while clearly better from a fitness standpoint, is also more risky.  Should your partner hunt hare, you end up with nothing. The result is that in evolutionary scenarios, it is often the case that populations are more likely to move from stag hunting to hare hunting than the reverse \citep{skyrms2004stag}.  

As with the prisoner's dilemma, though, mechanisms that correlate interaction among stag hunters can lead to the emergence and stabilization of prosocial behavior.  There are a few ways this can happen.  If individuals can choose their partners, stag hunters will tend to pick each other, leading to high payoffs for stag hunting. If individuals are able to communicate with each other, they can use even very simple signals to coordinate on stag hunting.  And in populations with a network structure, so that individuals tend to keep meeting the same neighbors for interaction, stag hunting can emerge spontaneously \citep{skyrms2004stag,alexander2009structural}. 

What can the stag hunt tell us about moral psychology?  This particular model seems especially useful in thinking about social trust.  An individual in a cooperative group will do poorly if they are too worried about social risks (and thus hunt for hare while the rest hunt stag).  The stag hunt may also tell us something about emotions like guilt or shame.  Individuals may be temporarily tempted to shirk social duties for their own benefit.  If this leads to poor payoffs in the end, emotions like guilt that decrease the chances of such shirking are directly beneficial \citep{o2016evolution}.

\subsection{Justice and Bargaining Games}

The two models thus far examine scenarios where actors can choose prosocial or antisocial types of behavior.  Let us now consider a set of models that are less commonly drawn upon in thinking about moral psychology---bargaining games.  These are games that capture scenarios where humans must divide some resource among themselves.  Note that these scenarios are ubiquitous.  Whenever social groups obtain resources, including food, tools, or building materials, they must decide how to divide them.  In addition, joint production involves another ubiquitous sort of bargaining---individuals must decide who will do how much of the work involved.

\begin{figure}
\centering
\includegraphics[width=.7\textwidth]{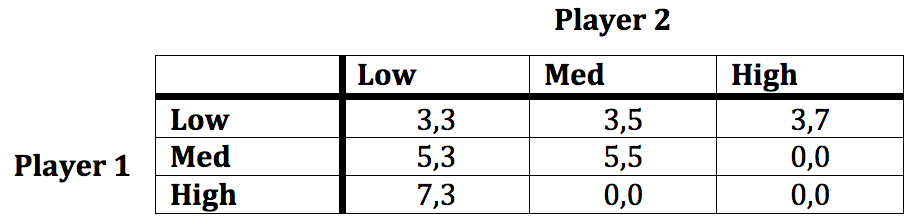}
\caption{A payoff table of the Nash demand game.  There are two players, each of whom choose one of three bargaining demands.  Payoffs are listed with player 1 first.
\label{fig:NDG}}
\end{figure}

There are several games used to represent such cases.  I will mention two, each of which correspond to different assumptions about control of the resource.  In the \emph{Nash bargaining game}, two actors make demands for a portion of a resource.  If these demands are compatible, they each get what they wanted.  If they are too aggressive and over-demand the resource, though, each gets a poor payoff.\footnote{This game was introduced by \citet{nash1950bargaining}.  It is sometimes called `divide the dollar', `divide the pie', or just the bargaining game.}  For simplicity sake, assume a resource of size 10, where each individual can make demands of 3, 5, or 7.  Figure \ref{fig:NDG} shows the payoff table for this game.  As is evident, when the payoffs sum to 10 or less, each actor gets what they demanded, otherwise they get 0.

There are three Nash equilibria of this game.  These are the three strategy pairings where the resource of 10 is exactly divided---the first actor demands 3 and the other 7, both demand 5, or the first actor demands 7 and the other 3.  One of the central evolutionary findings in this model is that the `fair' demand, where each player demands exactly half the resource, is more likely to evolve than populations with the other two unequal demands \citep{young1993evolutionary,skyrms2004stag}.\footnote{Though, see \citet{oconnorineq2018,d1998game}.}  This has been taken to help explain justice---both the cultural practice, and inequity aversion in humans \citep{skyrms2004stag, binmore2005natural,alexander2009structural}. I.e., if there is some evolutionary push towards fair demands, this might inform psychology related to equity.

The ultimatum game is a variation on the Nash bargaining game that yields very different predictions.  In this game, one actor controls the resource, and choose how much of it to offer to a partner.  The partner's choices are then to either accept what they are offered or to reject.  If they reject, neither partner gets anything.\footnote{This model was introduced by \citet{guth1982experimental}.}  The rational choice prediction is that the first player should offer as little as possible, on the expectation that the second player will prefer any offer to nothing, and thus accept.  In experimental set-ups, in fact, individuals make fairly high offers, and reject offers that are too low, though the details vary cross culturally \citep{guth1982experimental,henrich2006costly}.  This amounts to a kind of costly punishment---the second player in the game is willing to lower her payoff in order to lower the payoff of the first player.

Evolutionary models of the ultimatum bargaining game can help explain this seemingly irrational behavior.  Populations can evolve where actors make high offers, and reject low ones \citep{gale1995learning,harms1997evolution,skyrms2014evolution}.  These models may be informative in understanding the moral psychology of fairness, as well as retribution.

\subsection{Coordination and Norms}

Ethical systems tend to display in-group similarity and between-group variability.  For instance, as just mentioned, there is cross-cultural variation in how individuals play the ultimatum game, but relatively less variation within each culture \citep{henrich2006costly}.  It is even the case that ethical behaviors one culture finds abhorrent---infanticide, out-group homicide, honor killings---will be the norm elsewhere \citep{joyce2007evolution}.

\begin{figure}
\centering
\includegraphics[width=.6\textwidth]{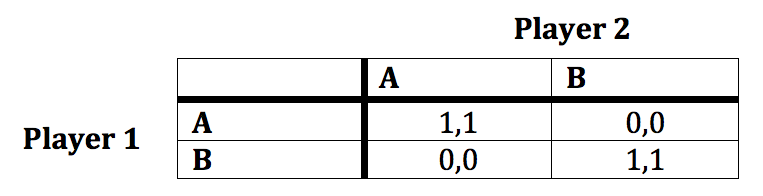}
\caption{A payoff table of a simple coordination game.  There are two players, each of whom chooses A or B.  Payoffs are listed with player 1 first.
\label{fig:coordination}}
\end{figure}

I will describe one last sort of model in this section---coordination games.  These are games where individuals choose one of two actions, and where their ultimate goal is to coordinate.  A classic example involves choosing which side of the road to drive on.  Drivers can choose the left or the right.  They generally care much less about which choice they make, and much more about making the same choice as other drivers.  Figure \ref{fig:coordination} shows the simplest possible coordination game, where actors get a payoff of 1 for choosing the same acts and 0 for choosing different ones.  The two Nash equilibria of this game are (unsurprisingly) the strategies where both actors do the same thing.

The need to coordinate behavior is ubiquitous in human groups.  In the moral sphere, there are serious problems for groups who do not share expectations about which behaviors are required and which forbidden \citep{stanford2018difference}.  The coordination game may be a helpful model in thinking about the aspects of norm psychology that get groups of humans behaving in coordinated ways, including aspects of social learning like conformity bias.

Before finishing this section, I would like to flag for the reader that the strategic situations outlined here are surely not the only ones that have been relevant to human social lives.  There are other models that might be used to help inform the evolution of moral psychology.  In addition, there are aspects of our moral psychology that have emerged in response to problems and opportunities that are not well modeled by games.  To give an example, most cultures have incest taboos.  These are culturally shaped, but humans also have a psychological tendency---dubbed the Westermark effect---to avoid those we grew up with as romantic partners \citep{shepher1983incest}.  This tendency likely evolved because incest leads to issues with deleterious gene mutations.  In other words, this moral psychological trait solves a social problem, but not a strategic one.

\section{Conclusion}

Section \ref{sec:Guilt} gave an example of how the sort of methodology I advocate here can work.  It starts by carefully identifying a moral psychological trait that might benefit from evolutionary explanation.  It next proceeds by drawing upon empirical literature to determine which behaviors are associated with this trait.  One can then use evolutionary models, perhaps drawing on the ones described in section \ref{sec:Models}, to develop a clear understanding of what leads to the selection of such behaviors, and thus what environments might lead to the selection of the psychological trait associated with that behavior.  Next one can use empirical literature to assess the success of the potential explanation developed.  In the case of guilt, as we saw, this sort of validation led us to downplay the importance of the \citet{frank1988passions} explanation of guilt, and put more weight on models of costly apology.

Of course, this sort of process will not work for all moral psychological traits.  These traits, as we saw in section \ref{sec:Methods}, are highly diverse.  The evolution of the psychology behind social imitation for example requires a very different sort of explanation than the evolution of guilt.  Furthermore, as noted, some moral psychology did not evolve to regulate strategic scenarios, and so may require a very different sort of methodology.  The goal here is not to be dogmatic.  Rather the goal is to give some good guidelines for further work on the evolution of moral psychology that makes use of the best epistemic tools available.

\section*{Acknowledgments}
Many thanks to Kyle Stanford for feedback on an earlier version of this manuscript.  Thanks to the editors of this volume, and an anonymous referee for their work and feedback.

\bibliographystyle{mla}
\bibliography{EvoMoral}
\end{document}